\documentclass{article}

\usepackage[final]{neurips_2023}

\usepackage[utf8]{inputenc} 
\usepackage[T1]{fontenc}    
\PassOptionsToPackage{hyphens}{url}
\usepackage{hyperref}       
\usepackage{url}            
\usepackage{booktabs}       
\usepackage{amsfonts}       
\usepackage{nicefrac}       
\usepackage{microtype}      
\usepackage{xcolor}         
\usepackage{graphicx}

\defcitealias{gplv3}{GPLv3}
\defcitealias{apachev2}{Apache-2.0}

\title{Towards Responsible Governance of Biological Design Tools}

\author{%
  Richard Moulange*\\
  Centre for the Governance of AI\\
  \texttt{rjm246@cam.ac.uk}
  \And
  Max Langenkamp*\\
  SecureDNA\\
  \texttt{maxlangenkamp@securedna.org}
  \And
  Tessa Alexanian*\\
  The Council on Strategic Risks\\
  \texttt{hello@tessa.fyi}
  \And
  Samuel Curtis*\\
  The Future Society\\
  \texttt{me@samuelmcurtis.com}
  \And
  Morgan Livingston\\
  Centre for the Governance of AI\\
  \texttt{livingston.a.morgan@gmail.com}
}

\begin{document}

\maketitle

\begin{abstract}
    Recent advancements in generative machine learning have enabled rapid progress in biological design tools (BDTs) such as protein structure and sequence prediction models. The unprecedented predictive accuracy and novel design capabilities of BDTs present new and significant dual-use risks. For example, their predictive accuracy allows biological agents, whether vaccines or pathogens, to be developed more quickly, while the design capabilities could be used to discover drugs or evade DNA screening techniques. Similar to other dual-use AI systems, BDTs present a wicked problem: how can regulators uphold public safety without stifling innovation? We highlight how current regulatory proposals that are primarily tailored toward large language models may be less effective for BDTs, which require fewer computational resources to train and are often developed in an open-source manner. We propose a range of measures to mitigate the risk that BDTs are misused, across the areas of responsible development, risk assessment, transparency, access management, cybersecurity, and investing in resilience. Implementing such measures will require close coordination between developers and governments.\let\thefootnote\relax\footnotetext{*equal contribution; order randomized}
\end{abstract}

\section{Introduction}

This paper outlines the distinct regulatory challenges posed by biological design tools (BDTs) and provides a range of governance measures to address the risks of their misuse.

We describe how the unprecedented predictive accuracy and novel design capabilities of BDTs have the potential to be misused by malicious actors. We then identify four key challenges in regulating BDTs: disagreement on permissible capabilities,  ineffectiveness of compute restriction on such small models, difficulty regulating decentralized and non-commercial development, and the potential for safeguards to be circumvented by open-sourcing model weights.

Preventing the misuse of biological design tools, while preserving their beneficial scientific uses, will require action at many phases of their lifecycle. We present a menu of 25 measures for mitigating risks of BDT misuse, including those that may be enacted by governments, by developers, and those which rely upon action by both, that we believe merit further exploration. We categorize these measures into six thematic areas: responsible development, risk assessment, transparency, access management, cybersecurity, and investing in resilience.

Discussions among the authors led us to highlight a subset of seven measures we believe to be particularly promising for reducing the risks posed by BDT misuse. These are: These are: model evaluations for dangerous capabilities, vulnerability reporting, structured access. know your customer, nucleic acid synthesis screening, securing lab equipment, and model-sharing infrastructure.

\section{Biological design tools present dual-use risks}
\label{BDT_dual_use}

Biological design tools (BDTs) are defined in \citet{sandbrink2023artificial} as ``systems that are trained on biological data and can help design new proteins or other biological agents''. These include protein folding tools, such as AlphaFold2 \citep{jumper2021highly}, protein design tools such as RFdiffusion \citep{watson2023novo}, genetic modification tools such as AlphaMissense \citep{cheng2023accurate} and many other others (see \citet{Nelson_Rose} for proposed subcategorization).

Advances in BDTs are rapidly reducing the gap between \textit{in silico} prediction and \textit{in vivo} performance, contributing to progress in many areas of biomedicine, including vaccine development, cancer therapy, and precision medicine \citep{kuhlman2019advances}. There are already several AI-designed molecules in early-stage clinical trials \citep{arnoldinside}, and AI tools may be able to speed up drug development, reduce costs, and increase the novelty of both new therapeutics and their molecular targets \citep{jayatunga2022ai}.

Unfortunately, many tools that enable high-precision design or prediction of biological molecules for scientific discovery or therapeutic development can be equally used for harm. Just as a protein-folding model facilitates the design of new drugs, it can also help malicious actors bypass software meant to screen DNA orders for potentially pathogenic sequences.
We highlight two significant dual-use characteristics of BDTs:
\vspace{-6pt}
\begin{itemize}
    \item[--] \textbf{Unprecedented predictive accuracy} reduces the amount of time, resources, and expertise required for experimental testing, iteration, and validation. This accelerates biological engineering, and biosecurity researchers have highlighted how this is likely to effectively shorten the risk chain for biological weapon development \citep{Nelson_Rose}.
    \vspace{-6pt}
    \item[--] \textbf{Novel design capabilities} enable the creation of pathogens that are more transmissible or virulent than known agents, that can evade current screening techniques, or that have the ability to target only specific species or sub-populations. BDTs may therefore may raise the ceiling of harm posed by a particular biological agent. In the chemical weapons context, the combination of a generative chemical model and a toxicity predictor tool was able to rediscover known nerve agents and suggest novel compounds that were predicted to be equally toxic \citep{urbina2022dual}. 
\end{itemize}
\vspace{-6pt}

There have been no publicly recorded attempts of BDTs being used to produce biological weapons or otherwise cause harm. However, malicious actors may plausibly seek to misuse these tools. The historical record shows that both state \citep{metcalfe2002short} and non-state \citep{danzig2011aum} actors have pursued the capability to cause large-scale harm through biological agents. At present, there are significant technical difficulties involved in manufacturing and delivering harmful biological agents, and \textbf{the authors expect the risk of misuse by BDTs to arise first and foremost from well-resourced and technically-competent actors}, such as those associated with state-sponsored bioweapons programs  (see \citet{yassif2023guarding} for an outline of such programs) or in existing research facilities.

As AI and bioengineering technologies mature, the pool of actors capable of misusing BDTs could broaden. In particular, we expect that, as with many types of scientific technology, newer versions of tools will be more user-friendly, and, as tools are integrated with one another, it will be easier for less technically-skilled actors to carry out complex biological workflows. Moreover, advances in LLMs may further reduce technical barriers. \citet{sandbrink2023artificial} speculates that Aum Shinrikyo may have been able to successively weaponize anthrax if their scientists had LLM-powered lab assistants to troubleshoot the conversion of a vaccine strain into its pathogenic form, and a recent US Executive Order on Safe AI \citep{whitehouseEO} warned about dual-use foundation models "lowering the barrier to entry for the development, acquisition, and use of biological weapons by non-state actors".

\section{Challenges in reducing misuse risk of biological design tools}
\label{challenges}

Regulators have increasingly expressed interest in mitigating the risks at the intersection of AI and biology. The Executive Order on Safe AI expressed concern that AI will “substantially [lower] the barrier of entry for non-experts to design… biological, radiological, or nuclear (CBRN) weapons” \citep{whitehouseEO}. U.S. Congress has also considered an ‘Artificial Intelligence and Biosecurity Risk Assessment Act’ designed specifically to begin to address the concerns outlined above \citep{markeybuddannouncement}. Policy conversations related to AI-enabled biological misuse are moving quickly, and it is an urgent matter to design proportionate policies that address misuse risks without unduly stifling innovation. 

Many governments are presently considering how to regulate AI systems, with a particular focus on foundation models \citep{anderljung2023frontier} and `high-risk’ systems, such as those used for biometric identification or for the operation and management of critical infrastructure \citep{EUParl}. Throughout these deliberations, a number of promising regulatory approaches have emerged, such as monitoring and restricting access to specific computational resources (“compute”) or introducing specific liability regimes for AI \citep{sullivan2019current, soyer2022artificial}. We find, however, that these regulatory approaches, which might be suitable for foundation models, will likely not translate well to BDTs. 


\subsection{Differentiating between harmful and permissible BDTs}
\label{distinguish}
As noted in Section 2, BDTs are dual-use technologies, with already-realized biomedical benefits and a simultaneous potential for harm. It is in our collective interest to avoid slowing down beneficial science by imposing disproportionate regulatory burdens on tools that pose little misuse risk.

A principal challenge in reducing misuse risk of BDTs will be understanding which properties and functions of BDTs correlate to risk. To that end, in a report on AI-facilitated biological weapon development, \citet{Nelson_Rose} propose a useful taxonomy of different BDT capabilities, including an assessment of the relative maturity of each capability, but do not suggest which tools are of particular concern. The Biden Administration's Executive Order on the Safe, Secure, and Trustworthy Development and Use of Artificial Intelligence includes an initiative to “evaluate the potential for AI to be misused to enable the development or production of CBRN threat” \citep{whitehouseEO}.

U.S. policy on dual-use research of concern articulates a set of “experiments of concern”, such as any which “alters the host range or tropism of the agent or toxin” \citep{durc}, , and provides a model for a technology-agnostic definition of high-risk activities; however, these policies are contested and under revision \citep{pannu}. Precise proposals for definitions of a “biological design tool of concern” will help developers and governments reach a consensus on which BDTs may possess misuse potential.

\subsection{Controlling access to compute is unlikely to be effective for many BDTs}
\label{compute}
Controlling access to computing infrastructure has attracted interest in the context of dual-use foundation models, which require large amounts of very specialized compute (i.e. the most advanced AI accelerator chips) to train. Training compute utilization has been shown to correlate with model performance and the “emergence” of new and more powerful capabilities \citep{kaplan2020scaling, ganguli2022predictability, wei2022emergent}. Thus, introducing controls on the most advanced AI accelerator chips and monitoring large training runs on compute clusters could curb the development of models harboring capabilities that pose extreme risks \citep{shavit2023does}. This careful delineation also helps avoid introducing burdensome regulation on less well-resourced actors. As a result, regulatory frameworks being developed in the EU, the UK, and the US reflect the critical role of computational resources in the development of frontier language models.\footnote{The European Parliament’s negotiating position on the proposed EU AI Act authorizes a European AI Office to “issue and periodically update guidelines on the thresholds that qualify training a foundation model as a large training run” and to “record and monitor known instances of large training runs” (see Amendment 529 of \citep{EUParl}). See also Case Study 3.9 in the UK white paper ‘A pro-innovation approach to AI regulation’ \citep{OfficeforAI} and the US export controls on specific AI chips \citep{Shivakumar}.}

Unfortunately, while such approaches may address risks from frontier foundation models, they seem unlikely to translate successfully to most BDTs. Training runs of frontier language models require hundreds or even thousands of advanced computer chips. GPT-4---currently, the most capable LLM---is reported to have about 1.8 trillion parameters and to have cost about US\$63mn to train \citep{Schreiner}. Even state-of-the-art BDTs have far fewer parameters and require orders of magnitude less compute to train. For example, RoseTTAFold, a popular open-source protein-folding model, has 130 million parameters and requires four weeks of training time on eight V100 GPUs \citep{baek2021accurate}, using at most 3.7$\times$10$^{20}$ FLOP.\footnote{We use \href{https://epochai.org/blog/estimating-training-compute}{Epoch's training compute calculator} with conservative estimates (28 days, 8 GPUs, NVIDIA Tesla V100 SMX2, FP16 precision and 62\% utilization rate).} At the time of writing, this would cost about US\$10,000\footnote{On 19 September 2023, Google Cloud pricing quoted a month’s usage of a V100 GPU at \$1267.28. By using more advanced chips, actors may be able to reduce costs further.} and is three orders of magnitude below the parameter threshold of 10$^{23}$ for biological models specified in the recent Executive Order on Safe AI that triggers additional reporting requirements. These far lower compute and cost requirements make applying compute governance to BDTs an impractical choice for mitigating misuse risks. 

\subsection{The decentralized and non-commercial nature of BDT development makes it harder to target and enforce regulation}
\label{decentralized}
Unlike dual-use foundation models, a wide variety of BDTs are developed by a range of actors across the world, often in academic or start-up labs. This decentralization makes it much more difficult to implement regulatory proposals that may be appropriate when dealing with only a few large technology companies. In particular,
\vspace{-6pt}
\begin{itemize}
    \item[--] \textbf{It is more difficult to track compliance}: with more actors involved, a regulator would have to spend more time and money ensuring that model developers are following the rules. In contrast, with only a few companies capable of developing a foundation model at the level of GPT-4, it is straightforward for regulators to identify those companies whose models might pose risks.
    \item[--] \textbf{There is a greater risk of regulatory arbitrage}: with BDT development spread across the world, many different countries may need to harmonize their legislation to effectively mitigate risks. This is also true for non-biology AI, but the overwhelming concentration of foundation model development in the US has meant that it is less of an immediate concern.
    \item[--] \textbf{It is harder to consult all model developers to develop best practices}: in contrast with the White House Voluntary Commitments, which initially applied to only seven advanced AI companies \citep{whitehouseVoluntary}, developing best practices for BDT development is likely to be slower and more difficult. Academic conferences and institutional collaborations like RosettaCommons will be key fora for developing such best practices.
\end{itemize}
\vspace{-6pt}

Moreover, that many BDT developers lack the resources of large technology companies increases the likelihood that regulation has the unintended effect of stifling innovation. For example, BDT developers cannot as easily implement potentially crucial cybersecurity features, such as those recommended in Section \ref{cybersecurity}, both through lack of funding and expertise.

\subsection{Open-source models are well-suited for science, but challenging targets for regulation}

The strong norms around open-source\footnote{By ‘open-source', we mean model weights and accompanying code that have been freely published on the internet, typically under an open-source license.} development of BDTs complicate the challenge of minimizing the risks of BDT misuse. Free access to code, data, and methodologies is the engine of modern science. By allowing other scientists to reproduce and verify experiments, open-source models facilitate scientific progress. Furthermore, open-source software is used in the vast majority of all software and has been estimated to generate upwards of US\$60 billion dollars worldwide \citep{ghosh2007study, synopsysreport}.

However, there are certain cases when open-sourcing a software undermines other important social values, such as public safety \citep{ospaper}. Open-sourcing renders software highly vulnerable to modification, which may be great for development, but also allows safety features to be circumvented. For instance, the safety filter used in Stable Diffusion---a state-of-the-art image generation system---can be nullified by deleting one line of code.\footnote{We do not provide further technical details, as part of responsible disclosure. This claim is informed by discussion with several technical researchers. See \citet{rando2022red} for a description of other attacks.} This enables anyone to produce prohibited images, including those featuring nudity and ‘deepfakes’ of real-life people. In the case of BDTs, open-source weights undermine the implementation of potential safeguards, which we discuss in Section \ref{measures}. 

Export controls are a common approach for mitigating the potential misuse of dual-use technologies, but often do not apply to `fundamental research'. Existing U.S. export controls apply to many dual-use items, including firearms parts, biological materials, and certain forms of information. Through a combination of regulations from the U.S. Department of Commerce, the Department of State, and the Office of Foreign Assets Control, US-based publishers and distributors of dual-use items are generally required to obtain a license prior to dissemination \citep{zeitouni2020milling}. However, information and materials that are "regularly published" or are a result of "fundamental research" are exempted from export controls \citep{15-cfr-734-8}, but the definitions of "published" and "fundamental research" are unclear. Department of Commerce regulations explicitly exempt software that is posted on the internet publicly (with the exception of specified encryption and firearm production software) \citep{15-cfr-734-7}. The Department of State, by contrast, does not provide an exemption for information published on the internet \citep{22-cfr-120-34}. \textbf{In the absence of specific guidance from regulatory agencies, it is unclear how the web of regulations around export controls applies to BDTs.}

Another possible mechanism for reducing risks of BDT misuse is imposing legal liability for developers of certain BDTs. However, this approach is stymied by the predominant licensing norms in open-source software. Virtually all open-source licenses come with clauses that absolve the developers of liability for misuse \citepalias{gplv3, apachev2}. Any liability regime would have to reconcile this existing norm.

\section{Measures for responsible governance of biological design tools}
\label{measures}

Mitigating risks arising from misuse of biological design tools, while preserving their beneficial scientific uses, will require action across various stages in their lifecycle. In Figure 1, we outline 25 measures that span the following phases of BDT development and use:

\textbf{1 Dataset Collection and Processing}: BDTs cannot be trained without specialized biological datasets. The development of datasets suitable for training BDTs has seen recent investment from both philanthropists and industry\footnote{For example, the Open Datasets Initiative funded by Schmidt Futures, the partnership of A-Alpha Bio with Lawrence Livermore National Labortory and the partnership of Ginkgo Bioworks and Google Cloud.} and been discussed as a target for biosecurity interventions \citep{soice2023can}. The recent Executive Order on Safe AI \citep{whitehouseEO} calls for a study that “considers the national security implications of the use of data and datasets, especially those associated with pathogens and omics studies."

\textbf{2 Model Development and Training}: Developing BDTs requires training the neural network(s) core to their functionality, optimizing and compressing trained models, and developing APIs, visualizations, and other interfaces to facilitate interaction with the model.

\textbf{3 Model Deployment}: There are many ways to make a BDT available for widespread use, such as open-sourcing model weights, using a hosting platform like Hugging Face or ColabFold, creating an API by which users can query the model, or releasing a database of predictions such as the ESM Metagenomic Atlas \citep{lin2023evolutionary} and AlphaFold Protein Structure Database \citep{varadi2022alphafold}.

\textbf{4 Operation and Monitoring}: For BDTs to be misused, their designs must be synthesized into physical biological agents, then deliberately or accidentally released. Though a full discussion of laboratory biosafety and biosecurity is beyond the scope of this paper, we highlight several promising interventions at the “digital-to-physical frontier” \citep{NTI} at which digital BDT designs are converted into physical biological materials.

We selected BDT-relevant measures from two reports on mitigating risks at the convergence of AI and biotechnology \citep{NTI, helenabioAI}, as well as recent reports and recommendations on best practices for dual-use foundation models \citep{whitehouseEO, IFP_testimony, UK_Emerging, schuett2023towards}. Although we believe every measure listed in this paper merits investigation, we do not believe they are all equally promising.



A range of actors will play a role in preventing the misuse of BDTs. Some measures, such as export controls, expanded liability, and mandating nucleic acid synthesis screening, require government regulation. Other measures, such as model evaluations and structured access, are currently being implemented as part of voluntary governance in developer communities.

We do not provide a comprehensive solution to BDT governance; best practices in this area are still being developed, not all measures we recommend are compatible with open source model weights (Supplementary Table 1), and the efficacy of existing measures is heavily contested (see Section 3). Instead, we present a menu of 25 measures for mitigating risks of BDT misuse that we believe merit further exploration. We divide these measures into six thematic areas, and believe seven measures are particularly promising based upon their perceived feasibility to implement and potential impact in risk mitigation: \textbf{model evaluations for dangerous capabilities, vulnerability reporting, structured access. know your customer, nucleic acid synthesis screening, securing lab equipment, and model-sharing infrastructure}.

\begin{figure}[!h]
    \centering
    \includegraphics[width=5.5in]{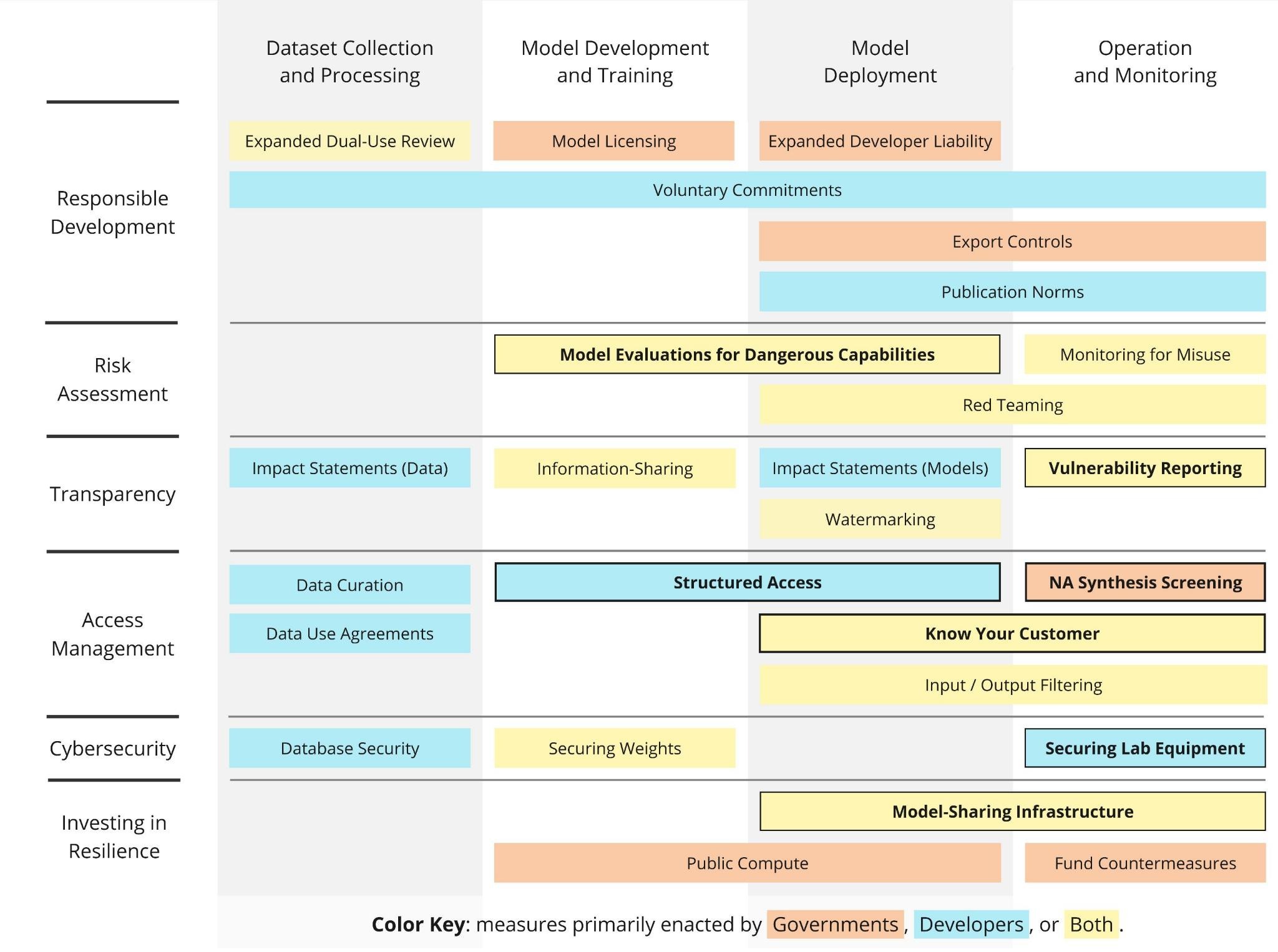}
    \caption{\textbf{Measures to mitigate misuse of biological design tools.} The measures are color-coded according to whether they primarily require action by governments (orange), by developers (blue), or both (yellow). Bolded measures are those the authors believe to be particularly promising for mitigating risks of BDT misuse.}
    \vspace{-10pt}
\end{figure}

\subsection{Responsible Development}
One way to prevent BDT misuse is to limit the development of BDTs that pose the greatest risks. As discussed in Section \ref{distinguish}, it is difficult to distinguish between harmful and permissible BDTs, so overall we argue that other parts of the development lifecycle are better targets for governance. That being said, we see six key opportunities for regulators to act to control the development of BDTs:

\textbf{1. Expanded Dual-Use Review}: Many countries have policies that regulate dual-use biological research, requiring institutional review, increased laboratory safety, and other precautions for research that could facilitate harm. Dual-use review could be extended to laboratory experiments that generate datasets which could be used to train BDTs with high misuse risk (e.g. measurements of cell toxicity, immune activation by viral vectors, or viral transmissibility).

\textbf{2. Model Licensing for Training or Release}: As is being considered for general-purpose AI systems, governments could consider requiring that BDT developers demonstrate compliance with certain institutional risk-mitigation processes or cybersecurity standards to obtain licenses to train or release certain high misuse risk BDT models.

\textbf{3. Expanded Developer Liability}: Expanded liability could incentivize developers to implement precautionary measures. The European Commission proposed an AI Liability Directive, which would extend consumer protections against damages caused with the involvement of AI systems \citep{Liabilitybriefing}. There is ambiguity about whether open-source software will fall under this directive, especially when used as a component within commercial products \citep{githubstatement}. One potential approach is to expand liability to developers who release model weights, but only if catastrophic damages occur \citep{gopal2023will}.

\textbf{4. Voluntary Commitments}: In a similar manner to the Voluntary Commitments by AI companies secured by the Biden Administration \citep{whitehouseEO}, leading BDT developers could voluntarily commit to undertake practices intended to reduce misuse risks. This could include, for instance, employing model evaluations prior to release, refraining from publishing preprints or releasing model weights until such evaluations are complete, post-deployment monitoring, or other measures listed in this paper. The Institute for Protein Design’s announcement of an effort to develop new voluntary guidelines suggest such commitments may be forthcoming \citep{ipd}.

\textbf{5. Export Controls}: As is imposed on types of information deemed dual-use, such as specifications for certain nuclear technologies, governments could impose export controls on certain BDTs deemed sufficiently dual-use, which would in effect restrict open-source release. However, presently, most software that is publicly available without restrictions upon further dissemination is exempt from US and EU regimes \citep{linuxfoundation}.

\textbf{6. Publication Norms}: Developers could adopt as a norm the practices of disclosing Impact Statements (Measure 11), withholding certain types of information from publications (or preprints) due to their potential for misuse, or refraining from releasing weights at the time of publishing until risk assessments or model evaluations have deemed them sufficiently safe to release. Likewise, journals and conferences could facilitate the adoption of norms by making publication or admission contingent on adherence to such practices.

\subsection{Risk Assessment}
Continual risk assessment across the development, deployment, and use of BDTs is necessary because their capabilities and potential for misuse are rapidly evolving. We separate risk assessment into three different measures, all of which are currently being tested by developers of general-purpose models:

\textbf{7. Model Evaluations for Dangerous Capabilities} \textit{(particularly promising)}: Benchmarks for dangerous capabilities could be designed by developers, independent third-parties, or government organizations \citep{shevlane2023model}. If dangerous capabilities are identified, developers can respond by pausing ongoing research and development, implementing Structured Access (Measure 17), or prioritizing cybersecurity (Measures 21-23).

\textbf{8. Red Teaming}: Structured tests could be conducted to identify flaws, limitations, and vulnerabilities in BDT software, in particular in the security features—if any—they possess, with the goal of remediation. This includes attempting to identify biological design tasks or input that elicit apparently harmful output, and efforts to circumvent restrictions to model weights, input/output screening techniques, and access controls.

\textbf{9. Monitoring for Misuse}: Developers could set up systems to automatically monitor user inputs and model outputs for potential misuse, with concerning results flagged for human review. This will help developers become aware of and respond to unanticipated potential for misuse, and should be paired with infrastructure to support Vulnerability Reporting (Measure 12). Monitoring is more easily done with API-based deployment, but local models could be bundled with oversight models that detect misuse. This is one of the emerging processes for frontier AI safety identified by the UK government \citep{UK_Emerging}.

\subsection{Transparency}
Transparency is needed for the development of safe, accountable, and trustworthy AI systems. The measures available to encourage transparency for BDTs seem much the same as those for other dual-use AI systems, including:

\textbf{10 Impact Statements}: Developers could establish as a norm that, upon releasing a model (or its constitutive components), they disclose why they believe doing so does not present a significant risk (or surpass some risk threshold). Standardized approaches for documenting the limitations and ethical considerations of models, such as model cards \citep{mitchell2019model}, could promote this norm.

\textbf{11 Information-Sharing with Regulators}: Model capabilities could be reported to regulators before (or shortly after) deployment. This requirement of foundation models has been proposed for the EU AI Act and may also apply to certain generative algorithms in China \citep{sheehan}.

\textbf{12 Vulnerability Reporting} \textit{(particularly promising)}: Developers could create (and regulators could require) mechanisms for software users to report concerning model behavior or other security flaws to their developers. This can help developers to quickly fix vulnerabilities, or, in extreme cases, rapidly roll back or withdraw a model.  As an incentive to report potential risks, “bug bounties” could provide financial rewards to the reporter. See \citet{whitehouseVoluntary, UK_Emerging, NTI}.

\textbf{13 Watermarking}: Leaving a recognizable signature in AI-generated biological designs could help with attribution and establishing liability for misuse. The White House Voluntary Commitments \citep{whitehouseVoluntary} and Chinese law \citep{sheehan} both include watermarking for generated content. Watermarking efforts could build upon existing research focused on genetic barcodes that track engineered microbes \citep{tellechealuzardo}.

\subsection{Access Management}
There is a spectrum of methods for releasing AI models, ranging from open-source releases to private, proprietary software \citep{solaiman2023gradient}. Different release methods give users different levels of access to AI system components. Developers will need to select access management strategies that appropriately trade-off between promoting innovation and preventing misuse. Some strategies that can be employed to limit access to dangerous capabilities of open-source BDTs are:

\textbf{14 Data Curation}: Specialized datasets used to train BDTs could be curated to limit  models’ ability to provide high-risk outputs. Curation may involve omitting a limited subset of research (e.g. on enhancement of potential pandemic pathogens) from public datasets \citep{soice2023can} or selectively adding noise to datasets \citep{campbell2023censoring} (e.g. adding noise to data related to nerve agents). 

\textbf{15 Data Use Agreements}: Users could be required to sign agreements outlining acceptable use, BDTs prior to obtaining access to datasets that could be used to train BDTs. Some biological databases, such as those containing patient data or infectious disease sequences, already require such agreements \citep{smith2022biosecurity}. Data use agreements outline circumstances when access can be revoked and provide legal recourse for dataset developers if they find their data has been misused.

\textbf{16. Structured access} \textit{(particularly promising)}: Structured access involves implementing role-based access control to software—generally in adherence to the principle of least privilege \citep{shevlane2022structured}. The archetypical method for implementing structured access is to deploy a model to the cloud and use an API to limit access to authenticated users. This kind of deployment limits the ability to circumvent Input/Output Filtering (Measure 20), facilitates the implementation of Know Your Customer (Measure 18) processes, provides opportunities for Monitoring Misuse (Measure 10), and supports Securing Weights (Measure 20), and allows access to be revoked for users exhibiting concerning (mis)use. Structured access is not compatible with open-source model weights, and cloud hosting expenses can be prohibitive for smaller non-commercial developers. Developers will need financial and technical support to implement structured access at scale, which could include Model-Sharing Infrastructure (Measure 23) and Public Compute (Measure 24).

\textbf{17. Know Your Customer} \textit{(particularly promising)}: Know Your Customer (KYC) processes involve collecting data to verify the legitimacy of a customer (or user) before providing them with a service. For example, KYC measures are required in the financial sector as part of anti-money laundering efforts. KYC could allow instances of misuse to be attributed to specific users, and, when Structured Access (Measure 16) is in place, access could be restricted (to some extent) for users with potentially concerning attributes. For datasets and models that are not fully open-sourced, KYC due diligence could be conducted before approving a data use agreement or granting model access.

\textbf{18. Nucleic Acid Synthesis Screening} \textit{(particularly promising)}: Preventing ordinary people from ordering smallpox DNA, as well as defending against more complex threats related to commercial nucleic acid synthesis, has been a longstanding priority for the biosecurity community \citep{williams} and has been proposed as a way to secure the “digital-to-physical frontier” \citep{helenabioAI} and protect against AI being used to engineer dangerous biological materials \citep{whitehouseVoluntary}. Nucleic acid synthesis screening could be universalized via regulation requiring synthesis companies to demonstrate compliance with screening standards, or by commitments by scientists not to purchase services unless companies demonstrate adherence to best practices in screening.

\textbf{19. Input/Output Filtering}: Developers could implement input filters refusing potentially high-risk inputs or outputs that may be considered high-risk (e.g. exhibiting similarity to more toxic compounds), potentially informed by data acquired by Monitoring for Misuse (Measure 10). This is likely to be more difficult for de novo designs. These filters may also be useful to prevent model extraction attacks, which reconstruct a local copy of a model based only on its API responses to user queries have been demonstrated for both image classification and natural language models \citep{krishna2019thieves}. In the LLM context, Anthropic responded to bioweapons-focused red teaming by adding classifier-based filters on model outputs \citep{anthropic}.

\subsection{Cybersecurity}
\label{cybersecurity}
Even BDTs with excellent access management could be misused due to cybersecurity vulnerabilities. Cybersecurity requirements are proposed in the EU AI Act \citep{EUParl} and included in the White House Voluntary Commitments \citep{whitehouseVoluntary}, and we consider investment in cybersecurity across the BDT development lifecycle a high priority:

\textbf{20. Database Security}: Public biological databases are used to build nucleic acid synthesis screening tools and track clinical events. Recent tabletop exercises have explored how database poisoning attacks could be used to undermine medical countermeasures \citep{bioisac}; cybersecurity best practices could allow database maintainers to detect tampering and restore integrity.

\textbf{21. Securing Weights}: Weights for closed-source models could be stolen by malicious external actors or leaked by insiders within developer communities. Investing in cybersecurity and insider threat safeguards to protect unreleased model weights is included in the White House Voluntary Commitments \citep{whitehouseVoluntary}.

\textbf{22. Securing Lab Equipment} \textit{(particularly promising)}: Securing lab equipment could include practices like implementing screening on DNA benchtop synthesizers to prevent the production of specific pathogens \citep{NTIbench}, ensuring that contract research organizations sequence customer-provided samples before running them on their equipment \citep{soice2023can}, and mandating minimum network security standards for biofoundry facilities.

\subsection{Investing in Resilience}
Governments should consider how their resources can be used to support developers in adopting governance measures. Successful implementation of many of the measures above will rely on cloud hosting of BDTs combined with effective cybersecurity practices. Without supportive investment, these measures may be infeasible or unsustainable for smaller non-commercial developers. Strategic investments could support responsible governance across the BDT lifecycle and decrease potential harms, should a BDT be misused. 

\textbf{23. Public Compute}: Governments could procure and redistribute computational resources to researchers using BDTs for research pandemic prevention and response. The establishment of this infrastructure could even expand access to these research tools—in particular, to underserved communities—through fast-tracked grants. This infrastructure could also provide a means for governments to promote best practices: access to computational resources should be conditional on adherence to responsible development and risk assessment measures, such as Publication Norms (Measure 6), Vulnerability Reporting (Measure 13), and other practices covered in Voluntary Commitments (Measure 4).
    
\textbf{24. Model-Sharing Infrastructure} \textit{(particularly promising)}: Governments could bear the costs of establishing and maintaining infrastructure to support secure and equitable  sharing of BDTs. This could include secure hosting infrastructure that complies with relevant cybersecurity standards and implements fair policies for granting Structured Access (Measure 17) through an API. There may be other ways for governments to support responsible model-sharing, such as offering standard templates for Data Use Agreements (Measure 16), software packages that support Monitoring for Misuse (Measure 10), or sets of Model Evaluations (Measure 7) to inform model release decision-making.

\textbf{25. Fund Countermeasures}: Governments could invest in a range of promising countermeasures for reducing potential harm caused by misuse of BDTs. These include investing in BDTs that bolster emergency preparedness and response \cite{ipd}, development of detection and surveillance approaches that are robust to novel design capabilities \cite{helenabioAI}, and broad-spectrum pandemic preparedness technologies that will reduce the severity of any future biological threat, whether deliberate or natural \citep{apolloprogram}.

\section{Conclusion}
Responsible governance of biological design tools will require mitigating the dual-use risks posed by BDTs without stifling scientific innovation or sacrificing biomedical advances. A multifaceted approach will be needed, combining both government- and developer-led measures to mitigate risk.

The development of BDTs with unprecedented predictive accuracy and novel design capabilities enables new frontiers in biomedicine, but risks misuse by actors seeking to cause large-scale harm through biology. Several regulatory proposals intended to reduce the risks of frontier language models, such as restricting compute access and export controls, are unlikely to be as effective for BDTs due to the significantly smaller amounts of compute needed to train BDTs and the generally decentralized and non-commercial nature of their development.

We present a range of risk mitigation measures and highlight seven that we believe to be particularly promising: model evaluations for dangerous capabilities, vulnerability reporting, structured access. know your customer, nucleic acid synthesis screening, securing lab equipment, and model-sharing infrastructure. Fully open-sourcing model weights is incompatible with structured access and know your customer efforts, and would undermine many of the other risk mitigation measures identified.

The scope of our analysis is constrained by the quickly evolving nature of both BDTs and the regulatory landscape. We stress that our proposals are theoretical and could benefit from testing and validation (i.e. regulatory sandboxing) to ascertain their efficacy and feasibility. We recognize the need for subsequent, more granular studies to refine the proposed measures. As BDTs and their potential misuse risks continue to evolve, empirical studies and insight from other efforts to regulate AI systems will be critical to refine and adapt policies. It is our hope that initiating these conversations now equips stakeholders to implement prudent, proportionate safeguards as these tools evolve.

\newpage

\section*{Acknowledgements}
We gratefully acknowledge Sophie Rose, Cassidy Nelson, and James Smith, as well as both anonymous reviewers from the NeurIPS 2023 Workshop on Regulatable ML, for helpful commentary on this manuscript. Any mistakes remain our own.

\bibliography{NeurIPS_23_BDTs}

\end{document}